\documentclass[11pt,fleqn,a4paper]{article}
\usepackage{amssymb,latexsym,amsmath,amsfonts}
\usepackage{graphicx}

    \advance\textheight by \topskip

    \newcommand\PVint{\mathop{\setbox0\hbox{$\displaystyle\intop$}%
    \hskip0.2\wd0%
    \vcenter{\hrule width0.6\wd0height0.5pt depth0.5pt}%
    \hskip-0.8\wd0%
    }\mskip-\thinmuskip\intop\nolimits}

    \def\tr{{\rm tr \,}}
    
    \def\Im{{\rm Im \,}}
    \def\norm1{\|_{\raisebox{-.5ex}{\scriptsize{\!1}}}}

    \newtheorem{theorem}{Theorem}[section]
    \newtheorem{lemma}[theorem]{Lemma}
    \newtheorem{corollary}[theorem]{Corollary}

    \newtheorem{Definition}[theorem]{Definition}
    
    \newtheorem{Remark}[theorem]{Remark}
    \newenvironment{remark}{\begin{Remark}\rm}{\end{Remark}}
    \newtheorem{Example}[theorem]{Example}

    \newenvironment{proof}%
    {\rm \trivlist \item[\hskip \labelsep{\bf Proof. }]}%
    {\hspace*{\fill}$\Box$\endtrivlist}
    \newenvironment{varproof}%
    {\rm \trivlist \item[\hskip \labelsep{\bf Proof}]}%
    {\hspace*{\fill}$\Box$\endtrivlist}

\begin{document}

    \begin{center} \Large\bf
        Universality for eigenvalue correlations from the
        modified Jacobi unitary ensemble
    \end{center}

    \

    \begin{center} \large
        A.B.J. Kuijlaars\footnote{Supported by FWO research project G.0176.02 and
        by INTAS project 00-272} \\
        \normalsize \em
        Department of Mathematics, Katholieke Universiteit Leuven, \\
        Celestijnenlaan 200 B,  3001 Leuven, Belgium \\
        \rm arno@wis.kuleuven.ac.be \\[3ex]
        \rm and \\[3ex]
        \large
        M. Vanlessen\footnote{Research Assistant of the Fund for Scientific Research -- Flanders (Belgium)}\\
        \normalsize \em
        Department of Mathematics, Katholieke Universiteit Leuven, \\
        Celestijnenlaan 200 B, 3001 Leuven, Belgium \\
        \rm maarten.vanlessen@wis.kuleuven.ac.be
    \end{center}\ \\[1ex]

\begin{abstract}
    The eigenvalue correlations of random matrices from the Jacobi Unitary Ensemble
    have a known asymptotic behavior as their size tends to infinity.
    In the bulk of the spectrum the behavior is described
    in terms of the sine kernel, and at the edge in terms of the Bessel kernel.
    We will prove that this behavior persists for the Modified Jacobi Unitary Ensemble.
    This generalization of the Jacobi Unitary Ensemble is associated with the modified
    Jacobi weight $w(x)=(1-x)^\alpha (1+x)^\beta h(x)$ where the extra factor $h$
    is assumed to be real analytic and strictly positive on $[-1,1]$.
    We use the connection with the orthogonal polynomials with respect to the modified
    Jacobi weight, and recent results on strong asymptotics derived by
    K.T-R McLaughlin, W.\ Van Assche and the authors.
\end{abstract}

\section{Introduction}
\setcounter{equation}{0}

In the early sixties, Dyson predicted that the local correlations between the eigenvalues of
ensembles of random matrices, when their size tends to infinity, have universal behavior in the bulk
of the spectrum. He expected that this universal behavior depends only on the type of the ensemble:
orthogonal, unitary or symplectic. This constitutes the famous conjecture of universality in the
theory of random matrices. For the classical ensembles (Hermite, Laguerre and Jacobi), this
conjecture has been proven, see for example
\cite{Mehta,NagaoSlevin,NagaoWadati1,ShiroNagaoWadati}.
For the unitary ensembles much more is known due to the connection with orthogonal
polynomials, and the universality conjecture in the bulk of the spectrum is
proved for a wide class of unitary ensembles,
see \cite{BleherIts,Deift,DKMVZ1,PasturShcherbina}.

At the edge of the spectrum this universal behavior breaks down.
For Hermite ensembles, it is known that the local correlations (at the soft edge)
can be expressed in terms of Airy functions \cite{BleherIts,Forrester1,TracyWidom1},
and for Jacobi and Laguerre ensembles (at the hard edge) in terms of Bessel functions
\cite{Forrester1,NagaoForrester,NagaoWadati,TracyWidom}. For example, for the Jacobi Unitary
Ensemble
\begin{equation} \label{JUE}
    \frac{1}{\tilde{Z}_n}  e^{ \tr \log w(M)} dM,
\end{equation}
where $w(x) = (1-x)^{\alpha} (1+x)^{\beta}$ is the Jacobi weight, the eigenvalue
correlations near $1$ are expressed in terms of the Bessel kernel
\begin{equation} \label{Besselkern}
    \mathbb{J}_\alpha(u,v) =
        \frac{J_\alpha(\sqrt{u})\sqrt{v}J'_\alpha(\sqrt{v})-
            J_\alpha(\sqrt{v})\sqrt{u}J'_\alpha(\sqrt{u})}{2(u-v)}
\end{equation}
as $n \to \infty$. $J_{\alpha}$ is the usual Bessel function of the first kind and order $\alpha$.
The order agrees with the exponent of $1-x$ in the Jacobi weight.

Nagao and Wadati \cite[$\S 6$]{NagaoWadati} expect that a universality result persists
for more general Jacobi-like ensembles, in the sense that the local form
of the weight function near $1$ determines the  eigenvalue correlation near $1$.
It is the aim of this paper to prove this universal behavior for
a generalization of the Jacobi Unitary Ensemble, which we call the Modified
Jacobi Unitary Ensemble (MJUE). The MJUE is given by (\ref{JUE}) with
modified Jacobi weight
\begin{equation} \label{Definitiew}
    w(x) = (1-x)^{\alpha} (1+x)^{\beta} h(x), \qquad \mbox{for } x \in (-1,1),
\end{equation}
where $\alpha, \beta > -1$ and the extra factor $h$ is real analytic and strictly positive on
$[-1,1]$. The Modified Jacobi Ensemble is a probability measure on the space of $n \times n$
Hermitian matrices with all eigenvalues in $(-1,1)$. The MJUE gives rise to a probability density
function of the $n$ eigenvalues $x_1, x_2, \ldots, x_n$ given by
\begin{equation}
    P^{(n)}(x_1, x_2, \ldots, x_n) = \frac{1}{Z_n}\prod_{j=1}^n w(x_j)\prod_{i<j}|x_i-x_j|^2,
\end{equation}
with $ x_1, x_2, \ldots, x_n \in (-1,1)$ and $Z_n$ a normalizing constant.

\bigskip
Dyson \cite{Dyson} showed, see also \cite{Deift,Mehta}, that we can express the correlation
functions $\mathcal{R}_{n,m}$
\[
    \mathcal{R}_{n,m}(x_1,\ldots x_m) =
    \frac{n!}{(n-m)!}\underbrace{\int \ldots \int}_{n-m}P^{(n)}(x_1,\ldots ,x_m,x_{m+1},\ldots
    ,x_n)dx_{m+1}\ldots dx_n
\]
in terms of orthogonal polynomials. Denote the $n$th degree orthonormal polynomial with respect to
the modified Jacobi weight $w$ by $p_n(z) =p_n(z;w)= \gamma_n z^n+\cdots$, $\gamma_n > 0$. Then
$\mathcal{R}_{n,m}(x_1,\ldots x_m) = \det\left(K_n(x_i,x_j)\right)_{1\leq i,j \leq m}$, where
\begin{equation}\label{DefinitieKn}
    K_n(x,y) = \sqrt{w(x)}\sqrt{w(y)}\sum_{j=0}^{n-1}p_j(x) p_j(y).
\end{equation}
By the Christoffel-Darboux formula, we have
\begin{equation} \label{ChristoffelDarboux}
    K_n(x,y) = \sqrt{w(x)} \sqrt{w(y)} \, \frac{\gamma_{n-1}}{\gamma_n}\,
    \frac{p_n(x) p_{n-1}(y) - p_{n-1}(x) p_n(y)}{x-y},
\end{equation}
which shows that asymptotic properties of $K_n$ are
intimately related with asymptotics of the orthogonal polynomials
$p_n$ as $n \to \infty$.

In a previous paper with K.T-R McLaughlin and W.\ Van Assche \cite{KMVV}, we studied the
asymptotics of the polynomials that are orthogonal with respect to the modified Jacobi weight. We
used the Riemann-Hilbert formulation for orthogonal polynomials of Fokas, Its, and Kitaev
\cite{FokasItsKitaev} and the steepest descent method for Riemann-Hilbert problems of Deift and
Zhou \cite{DeiftZhou}. In \cite{KMVV} we concentrated on the asymptotics of the polynomials away
from the interval $[-1,1]$, but the Riemann-Hilbert method gives uniform asymptotics in all regions
in the complex plane. Here we are interested in the behavior on $[-1,1]$, and in particular near
the endpoints $\pm 1$. The Riemann-Hilbert method was applied before to orthogonal polynomials by
Deift and co-authors \cite{Deift,DKMVZ1,DKMVZ2,DKMVZ3,KriecherbauerMcLaughlin}. They studied
orthogonal polynomials on the real line with varying weights, and used the asymptotics to prove the
universality in the bulk of the spectrum for the associated unitary ensembles. We apply the same
method to prove the universality at the edge of the spectrum for the MJUE. Our main result is the
following.

\begin{theorem}\label{TheoremUniversalityEdge}
    Let $w$ be the modified Jacobi weight {\rm (\ref{Definitiew})} and
    let $K_n$ be the kernel {\rm (\ref{DefinitieKn})} associated with $w$.
    Then the following holds.
    \begin{enumerate}
    \item[\rm (a)]
        For $x \in (-1,1)$, we have as $n \to \infty$,
        \begin{equation} \label{DensityofStates}
            \frac{1}{n} K_n(x,x) = \frac{1}{\pi \sqrt{1-x^2}} + O\left(\frac{1}{n}\right).
        \end{equation}
        The error term is uniform for $x$ in compact subsets of $(-1,1)$.
    \item[\rm (b)]
        Let $\xi(x) = \frac{1}{\pi \sqrt{1-x^2}}$. Then
        for $x \in (-1,1)$ and $u, v \in \mathbb R$, we have as $n \to \infty$,
        \begin{equation} \label{universalbulk}
            \frac{1}{n \xi(x)} K_n\left(x + \frac{u}{n \xi(x)}, x + \frac{v}{n \xi(x)}\right)
            = \frac{\sin \pi (u-v)}{\pi(u-v)} + O\left(\frac{1}{n}\right).
        \end{equation}
        The error term is uniform for $x$ in compact subsets of $(-1,1)$ and for $u,v$ in
        compact subsets of $\mathbb R$.
    \item[\rm (c)]
        For $u,v \in (0,\infty)$, we have as $n \to \infty$,
        \begin{equation} \label{universalnear1}
            \frac{1}{2n^2}K_n\left(1-\frac{u}{2n^2},1-\frac{v}{2n^2}\right)
            = \mathbb{J}_\alpha(u,v)+O\left(\frac{u^\frac{\alpha}{2}v^\frac{\alpha}{2}}{n}\right),
        \end{equation}
        where $\mathbb J_{\alpha}$ is the Bessel kernel given by {\rm (\ref{Besselkern})}.
        The error term is uniform for $u,v$ in bounded subsets of $(0,\infty)$.
    \end{enumerate}
\end{theorem}
Note that the error term in (\ref{universalnear1})
holds uniformly for $u,v$ in {\em bounded} subsets of $(0,\infty)$,
not just in compact subsets.
By symmetry, there is a corresponding universality result near $-1$.

\bigskip

The eigenvalue density is the 1--point correlation function $\mathcal{R}_{n,1}(x)=K_n(x,x)$, see
for example \cite{Mehta}. Therefore, part (a) of the theorem yields the asymptotic eigenvalue
density $\mathcal{R}_{n,1}(x)\sim n\xi(x)$ as $n\to\infty$. This result is in agreement with
\cite{Leff,NagaoWadati1}. The scaling in (\ref{universalbulk})
has the effect that $x$ is the new origin and that the asymptotic eigenvalue density at $x$ is 1.
At the endpoints, (\ref{DensityofStates}) breaks down, and the eigenvalue density is $O(n^2)$ as
$n\to\infty$, near the endpoints, see for example \cite{Leff}. This explains the scaling in
(\ref{universalnear1}).

Part (b) of the theorem states the universality (independent of the choice of $\alpha,\beta, h$ and
$x$) for $K_n$ in the bulk of the spectrum. It extends the result of Nagao and Wadati
\cite[(4.19)]{NagaoWadati1} for the case that $h\equiv 1$. At the edge 1 of the spectrum we have a
universality class for $K_n$ (independent of the choice of $\beta$ and $h$) which is only affected
by the local form of the modified Jacobi weight near 1, see part (c).

\bigskip

Using Theorem \ref{TheoremUniversalityEdge} we can answer local statistical quantities concerning
the eigenvalues. Here we follow \cite{Deift,DKMVZ1}. The probability $P_n(a,b)$ that there are no
eigenvalues in the interval $(a,b)\subset (-1,1)$ is given by
\[
    P_n(a,b) = \det(I-K_n),
\]
where $K_n$ is the trace class operator with integral kernel $K_n(x,y)$ acting on $L^2(a,b)$, and
where $\det(I-K_n)$ is the Fredholm determinant. For a fixed interval $(a,b)$ we have that
$P_n(a,b)\to 0$, as $n\to\infty$. So, to understand the asymptotic behavior of $P_n$ at the edge of
the spectrum, we will look at intervals near the edges which shrink with $n$, and we are led to
consider the asymptotic behavior of $P_n\left(1-\frac{s}{2n^2},1\right)$ as $n\to\infty$, where
$s>0$. We have the following universality for $P_n$ at the edge 1 of the spectrum, depending on the
parameter $\alpha$ but independent of the choice of $\beta$ and $h$.
\begin{corollary}\label{Corollary2UniversalityEdge}
    For $s>0$, we have
    \begin{equation}
        \lim_{n \to \infty} P_n\left(1-\frac{s}{2n^2},1\right)
        = \det(I-\mathbb{J}_{\alpha,s}),
    \end{equation}
    where $\mathbb{J}_{\alpha,s}$ is the integral operator
    with kernel $\mathbb{J}_\alpha(u,v)$ acting on $L^2(0,s)$,
    and $\det(I - \mathbb{J}_{\alpha,s})$ is the Fredholm determinant.
\end{corollary}

As mentioned before, our main tool in proving Theorem 1.1 is the asymptotic analysis
of the Riemann-Hilbert problem for the orthogonal polynomials with respect to
the modified Jacobi weight, as  developed in \cite{KMVV}. We give an overview of this
work in section 2.  This approach is able to give strong and uniform asymptotics
for the orthogonal polynomials in every region in the complex plane,
which we also review in section 2.
The proofs of Theorem 1.1 and Corollary 1.2 are given in section 3.

\section{Riemann--Hilbert problem for orthogonal polynomials}
\setcounter{equation}{0}

In this section we will recall the Riemann--Hilbert problem (RH problem) from \cite{KMVV}
for the orthogonal polynomials for the modified Jacobi weight,
and the steps in the steepest descent method that
are used to obtain the asymptotics  on the interval.

The analysis starts from a characterization of the monic orthogonal polynomials $\pi_n$ for the
modified Jacobi weight $w$ given by (\ref{Definitiew}) as a solution of a RH problem for a $2\times
2$ matrix valued function $Y(z)=Y(z;n,w)$. This characterization of orthogonal polynomials is due
to Fokas, Its, and Kitaev \cite{FokasItsKitaev}. The conditions (\ref{RHPYd}) and (\ref{RHPYe}) are
needed to control the behavior near the endpoints, see \cite{KMVV} for discussion.
\begin{enumerate}
    \item[(a)]
        $Y(z)$ is  analytic for $z\in\mathbb C \setminus [-1,1]$.
    \item[(b)]
        $Y$ possesses continuous boundary values for $x \in (-1,1)$
        denoted by $Y_{+}(x)$ and $Y_{-}(x)$, where $Y_{+}(x)$ and $Y_{-}(x)$
        denote the limiting values of $Y(z)$ as $z$ approaches $x$ from
        above and below, respectively, and
        \begin{equation}\label{RHPYb}
            Y_+(x) = Y_-(x)
            \begin{pmatrix}
                1 & w(x) \\
                0 & 1
            \end{pmatrix},
            \qquad\mbox{for $x \in (-1,1)$.}
        \end{equation}
    \item[(c)]
        $Y(z)$ has the following asymptotic behavior at infinity:
        \begin{equation} \label{RHPYc}
            Y(z) = \left(I+ O \left( \frac{1}{z} \right)\right)
            \begin{pmatrix}
                z^{n} & 0 \\
                0 & z^{-n}
            \end{pmatrix}, \qquad \mbox{as $z\to\infty$.}
        \end{equation}
    \item[(d)]
        $Y(z)$ has the following behavior near $z=1$:
        \begin{equation}\label{RHPYd}
            Y(z) = \left\{
            \begin{array}{cl}
                O\begin{pmatrix}
                    1 & |z-1|^{\alpha} \\
                    1 & |z-1|^{\alpha}
                \end{pmatrix},
                &\mbox{if $\alpha<0$,} \\[2ex]
                O\begin{pmatrix}
                    1 & \log|z-1| \\
                    1 & \log|z-1|
                \end{pmatrix},
                &\mbox{if $\alpha=0$,} \\[2ex]
                O\begin{pmatrix}
                    1 & 1 \\
                    1 & 1
                \end{pmatrix},
                &\mbox{if $\alpha>0$,}
            \end{array}\right.
        \end{equation}
        as $z \to 1$, $z \in \mathbb C \setminus [-1,1]$.
    \item[(e)]
        $Y(z)$ has the following behavior near  $z=-1$:
        \begin{equation} \label{RHPYe}
            Y(z) = \left\{
            \begin{array}{cl}
                O\begin{pmatrix}
                    1 & |z+1|^{\beta} \\
                    1 & |z+1|^{\beta}
                \end{pmatrix}, &\mbox{if $\beta<0$,} \\[2ex]
                O\begin{pmatrix}
                    1 & \log|z+1| \\
                    1 & \log|z+1|
                \end{pmatrix},
                &\mbox{if $\beta=0$,} \\[2ex]
                O\begin{pmatrix}
                    1 & 1 \\
                    1 & 1
                \end{pmatrix},
                &\mbox{if $\beta>0$,}
            \end{array}\right.
        \end{equation}
        as $z \to -1$, $z \in \mathbb C\setminus [-1,1]$.
\end{enumerate}

The unique solution of this RH  Problem is given by
\begin{equation} \label{RHPYsolution}
    Y(z) =
    \begin{pmatrix}
        \pi_n(z) & \frac{1}{2\pi i} \int_{-1}^1  \frac{\pi_n(x) w(x)}{x-z}dx \\[2ex]
        -2\pi i \gamma_{n-1}^2 \pi_{n-1}(z) & -\gamma_{n-1}^2 \int_{-1}^1 \frac{\pi_{n-1}(x)w(x)}{x-z} dx
    \end{pmatrix},
\end{equation}
where $\pi_n$ is the monic polynomial of degree $n$ orthogonal with respect to the weight $w$ and
with $\gamma_n$ the leading coefficient of the orthonormal polynomial $p_n$.
\bigskip

We apply a number of transformations $Y\mapsto T\mapsto S\mapsto R$ to the original RH problem in
order to arrive at a RH problem for $R$, which is normalized at infinity, and whose jump matrices
are uniformly close to the identity matrix. Then, $R$ is uniformly close to the identity matrix,
and by tracing back the steps we deduce the asymptotic behavior of $Y$.

In the first transformation we turn the original RH problem into an equivalent RH problem for $T$,
which is normalized at infinity and with a jump matrix whose diagonal elements are oscillatory. Let
\[
    \sigma_3
    = \begin{pmatrix}
        1 & 0 \\
        0 & -1
    \end{pmatrix}
\]
be the Pauli matrix, and let $T$ be given by
\begin{equation}\label{YinT}
    T(z) = 2^{n\sigma_3}Y(z)\varphi(z)^{-n\sigma_3},
\end{equation}
where $\varphi(z)=z+(z^2-1)^{1/2}$ for $z\in\mathbb{C}\setminus [-1,1]$, so that $\varphi$ is the
conformal map from $\mathbb{C}\setminus[-1,1]$ onto the exterior of the unit circle. Then $T$
satisfies a RH problem which is normalized at infinity (i.e., $T(z) \to I$ as $z \to \infty$), and
\[
    T_+(x) = T_-(x)
    \begin{pmatrix}
        \varphi_+(x)^{-2n} & w(x) \\
        0 & \varphi_-(x)^{-2n}
    \end{pmatrix},
    \qquad \mbox{for } x \in (-1,1).
\]

\begin{figure}[ht]
    \center{\resizebox{8cm}{!}{\includegraphics{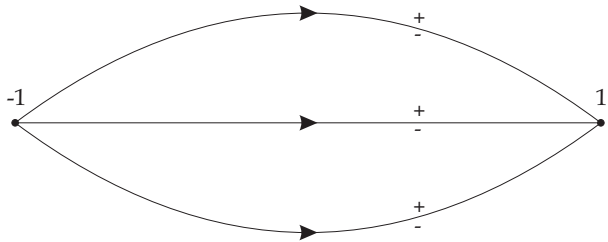}}
    \caption{The lens $\Sigma$}\label{figure1}}
\end{figure}
The jump matrix for $T$ factors into a product of three matrices. Using this factorization, we
transform the RH problem for $T$ into an equivalent RH problem for $S$, with jumps on a lens shaped
contour $\Sigma$, as in Figure \ref{figure1}. $S$ is defined in terms of $T$ by
\begin{equation} \label{TinS}
    S(z) =
    \left\{\begin{array}{cl}
        T(z), & \mbox{for $z$ outside the lens,} \\[2ex]
        T(z)
        \begin{pmatrix}
            1 & 0 \\
            -w(z)^{-1}\varphi(z)^{-2n} & 1
        \end{pmatrix}, & \mbox{for $z$ in the upper part of the lens,} \\[2ex]
        T(z)
        \begin{pmatrix}
            1 & 0 \\
            w(z)^{-1}\varphi(z)^{-2n} & 1
        \end{pmatrix}, & \mbox{for $z$ in the lower part of the lens.}
    \end{array}\right.
\end{equation}

For the third transformation, we need to construct parametrices
in the outside region and near the endpoints $\pm 1$.
Constructing the parametrix in the outside region, we need the
Szeg\H{o} function $D(z)=D(z;w)$
associated with the weight $w$ given by
\begin{equation} \label{Szegofunctie}
    D(z) = \exp\left( \frac{(z^{2}-1)^{1/2}}{2\pi}\int_{-1}^{1}
    \frac{\log w(x)}{\sqrt{1-x^{2}}} \frac{dx}{z-x}\right),\qquad\mbox{for }
    z\in\mathbb{C}\setminus[-1,1].
\end{equation}
The Szeg\H{o} function is a non-zero analytic function on $\mathbb C \setminus [-1,1]$
such that $D_+(x) D_-(x)  = w(x)$ for $x \in (-1,1)$.
The parametrix $N$ in the outside region is given by
\begin{equation}\label{RHPNsolution}
    N(z) = D_{\infty}^{\sigma_3}
    \begin{pmatrix}
        \frac{a(z) + a(z)^{-1}}{2} & \frac{a(z) - a(z)^{-1}}{2i} \\[1ex]
        \frac{a(z) - a(z)^{-1}}{-2i} & \frac{a(z) + a(z)^{-1}}{2}
    \end{pmatrix}
    D(z)^{-\sigma_3},
    \end{equation}
where $D_\infty = \lim_{z\to\infty}D(z)$ and  $a(z) = (z-1)^{1/4}(z+1)^{-1/4}$.

\begin{figure}[ht]
    \center{\resizebox{5cm}{!}{\includegraphics{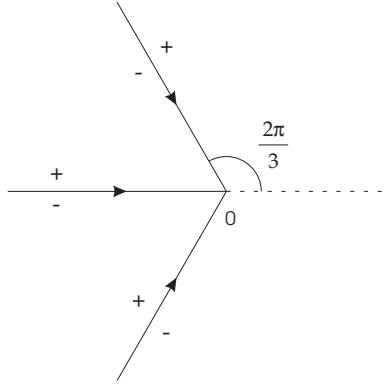}}
    \caption{The contour $\Sigma_{\Psi}$}\label{figure2}}
\end{figure}

Next, we define a parametrix $P$ in $U_\delta$, which is the disk with radius $\delta$ and
center $1$, where $\delta>0$ is sufficiently small, by
\begin{equation}\label{RHPPsolution}
    P(z) = E_{n}(z)\Psi(n^{2}f(z))W(z)^{-\sigma_{3}}\varphi(z)^{-n\sigma_{3}},
\end{equation}
where $f(z)$ and $W(z)$ are scalar functions given by
\begin{equation}\label{deff(z)}
    f(z) = \frac{1}{4}\bigl(\log \varphi(z)\bigr)^2,
\end{equation}
and
\begin{equation}\label{W^2(z)}
    W(z) = (z-1)^{\alpha/2}(z+1)^{\beta/2}h^{1/2}(z).
\end{equation}
In (\ref{RHPPsolution}) $\Psi(\zeta)$ is a $2\times 2$ matrix valued function defined for
$\zeta\in\mathbb{C}\setminus\Sigma_{\Psi}$, where $\Sigma_{\Psi}$ is the contour shown in Figure
\ref{figure2}.
For our purpose here, it suffices to know the expression of $\Psi(\zeta)$ for $2\pi/3 <
\arg \zeta < \pi$. This is given in terms of the Hankel functions $H_\alpha^{(1)}$
and $H_\alpha^{(2)}$. For $2\pi/3 <\arg \zeta <\pi$, we have
\begin{equation}\label{RHPPSIsolution2}
    \Psi(\zeta) =
    \begin{pmatrix}
        \frac{1}{2} H_{\alpha}^{(1)}(2(-\zeta)^{1/2}) &
        \frac{1}{2} H_{\alpha}^{(2)}(2(-\zeta)^{1/2}) \\[1ex]
        \pi \zeta^{1/2} \left(H_{\alpha}^{(1)}\right)'(2(-\zeta)^{1/2}) &
        \pi \zeta^{1/2} \left(H_{\alpha}^{(2)}\right)'(2(-\zeta)^{1/2})
    \end{pmatrix} e^{\frac{1}{2}\alpha\pi i \sigma_3 }.
\end{equation}
The factor $E_n(z)$ in (\ref{RHPPsolution}) is  given by
\begin{equation} \label{eq536}
    E_n(z) = N(z) W(z)^{\sigma_3} \frac{1}{\sqrt{2}}
    \begin{pmatrix}
        1 & -i \\
        -i & 1
    \end{pmatrix}
    f(z)^{\sigma_3/4} \left(2\pi n \right)^{\sigma_3/2}.
\end{equation}
$E_n(z)$ is analytic in a full neighborhood of $U_{\delta}$.

There is a similar definition for the parametrix $\tilde{P}$ in
a $\delta$ neighborhood $\tilde{U}_{\delta}$ of $-1$, see \cite{KMVV}
for details.

\begin{figure}[ht]
    \center{\resizebox{8cm}{!}{\includegraphics{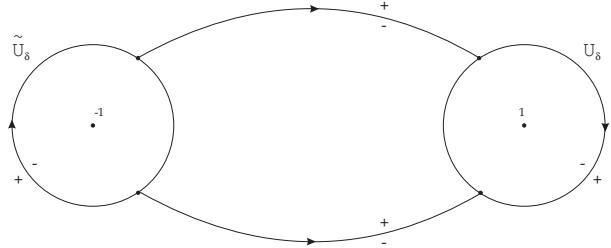}}
    \caption{System of contours $\Sigma_{R}$}\label{figure3}}
\end{figure}

We then have all the ingredients for the third transformation.
We define
    \begin{eqnarray}
        \label{DefRBuiten}
        R(z) &=&
            S(z) N^{-1}(z),\qquad \mbox{for }
            z\in\mathbb{C}\setminus(\overline{U}_{\delta}\cup\overline{\tilde{U}}_{\delta}
            \cup \Sigma), \\[1ex]
        \label{DefRU}
        R(z) &=&
            S(z) P^{-1}(z),\qquad \mbox{for } z \in U_{\delta} \setminus \Sigma, \\[1ex]
        \label{DefRUTilde}
        R(z) &=&
            S(z) \tilde{P}^{-1}(z),\qquad \mbox{for }
            z \in \tilde{U}_{\delta} \setminus\Sigma.
    \end{eqnarray}
Then $R$ satisfies a RH problem with jumps on the system of contours $\Sigma_R$ shown in Figure
\ref{figure3}.

Note that $R$ depends on $n$, but the contour $\Sigma_R$ does not depend on $n$. The jump matrices
for the RH problem for $R$ turn out to be uniformly close to the identity matrix with error term
$O(\frac{1}{n})$. Then it follows that
\begin{equation} \label{asymptoticsR}
    R(z) = I+O\left(\frac{1}{n}\right), \qquad \mbox{as } n\to\infty,
\end{equation}
uniformly for $z\in\mathbb{C}\setminus \Sigma_R$, and also,
see \cite[Section 8.1]{Deift} in particular formulas (8.19) and (8.20),
\begin{equation} \label{asymptoticsdR}
    \frac{d}{dz} R(z) = O\left(1 \right), \qquad \mbox{as } n \to\infty,
\end{equation}
uniformly for $z \in \mathbb C \setminus \Sigma_R$.
In the following we will also use that
\begin{equation} \label{determinantR}
    \det R(z) \equiv 1.
\end{equation}

\begin{remark}
The asymptotic result (\ref{asymptoticsR}) for $R$ may be refined
and it is possible to obtain a full asymptotic expansion
\[
    R(z) \sim I + \sum_{k=1}^{\infty} \frac{R_k(z)}{n^k}, \qquad \mbox{as } n \to \infty.
\]
The matrix coefficients $R_k(z)$ are explicitly computable
in function of $\alpha$, $\beta$, and the analytic factor $h$ in the
modified Jacobi weight. The first two coefficients are computed
in \cite{KMVV}.
\end{remark}

\begin{remark}
In \cite{KMVV} we also derived asymptotic expansions for the polynomials $p_n$ and $\pi_n$
uniformly valid in compact subsets of $\overline{\mathbb C} \setminus [-1,1]$, as well as
asymptotic expansions for the leading coefficient $\gamma_n$ and for the coefficients in the
recurrence relation satisfied by the orthonormal polynomials. For example, we have by \cite[Theorem
1.4]{KMVV} uniformly for $z \in \mathbb C \setminus [-1,1]$,
\[
    \frac{2^n \pi_n(z)}{\varphi(z)}
    = \frac{D_{\infty}}{D(z)} \frac{\varphi(z)^{1/2}}{\sqrt{2}(z^2-1)^{1/4}}
    \left[1 + O\left(\frac{1}{n}\right)\right],
    \qquad \mbox{as } n \to \infty.
\]
The $O(\frac{1}{n})$ term can be developed into a complete asymptotic
expansion in powers of $n^{-1}$.

The Riemann-Hilbert method also leads to strong asymptotics on the interval
$(-1,1)$ and near the endpoints $\pm 1$. While these results are closely
related to the asymptotics of $K_n$ as given in Theorem 1.1, we do not
actually rely on them in the proof of Theorem 1.1.
Therefore we  state here the asymptotics of the orthogonal polynomials
without proof. The proof is similar to the proof of Theorem 1.1, and
in fact somewhat simpler. See also \cite{DKMVZ1,DKMVZ2}.

As before, we let $\delta > 0$ be the radius of the disks $U_{\delta}$
and $\tilde{U}_{\delta}$. Then we have for $x \in (-1+\delta, 1-\delta)$,
\begin{eqnarray} \nonumber
    \pi_n(x) & = & \frac{\sqrt{2} D_{\infty}}{2^n \sqrt{w(x)}(1-x^2)^{1/4}}
    \left[ R_{11}(x) \cos\left( (n+1/2) \arccos x + \psi(x) - \frac{\pi}{4} \right) \right. \\
    & & \qquad \qquad \left.
        - \frac{i}{D_{\infty}^2} R_{12}(x) \cos\left((n-1/2)\arccos x + \psi(x) - \frac{\pi}{4}\right)
        \right], \label{asymptotiekpinbulk}
\end{eqnarray}
where $R_{11}(x) = 1 + O(\frac{1}{n})$ and $R_{12}(x) = O(\frac{1}{n})$ as $n \to \infty$,
uniformly for $x \in (-1+\delta, 1- \delta)$. The $O(\frac{1}{n})$ terms have a complete asymptotic
expansion, see \cite{KMVV}. The function $\psi(x)$ in (\ref{asymptotiekpinbulk}) is given by
\begin{eqnarray} \nonumber
    \psi(x) & = & \frac{\sqrt{1-x^2}}{2\pi}\PVint_{-1}^1\frac{\log w(t)}{\sqrt{1-t^2}}
        \frac{dt}{t-x} \\
    & = & \label{functiepsi}
     \frac{1}{2}\Bigl(\alpha(\arccos x-\pi)+\beta \arccos x\Bigr)
        +\frac{\sqrt{1-x^2}}{2\pi}\PVint_{-1}^1\frac{\log h(t)}{\sqrt{1-t^2}}
        \frac{dt}{t-x},
\end{eqnarray}
where the integral is a Cauchy principal value integral.
We remark that, under various assumptions, asymptotic results on the
interval of orthogonality have been established by many authors, see for example
\cite{DKMVZ1,DKMVZ2,LevinLubinsky,Lubinsky,Nevai,Szego,Totik}.

For $x \in (1-\delta, 1)$, the following is valid
\begin{eqnarray} \nonumber
    \pi_n(x) & = & \frac{\sqrt{\pi}D_{\infty}}{2^n \sqrt{w(x)}}
        \frac{(n \arccos x)^{1/2}}{(1-x^2)^{1/4}} \, \times \\[1ex]
        & & \nonumber
            \left[R_{11}(x) \left(\cos \zeta_1(x) J_{\alpha}(n \arccos x) +
                \sin \zeta_1(x) J_{\alpha}'(n \arccos x) \right) \right. \\
        & & \qquad \left. - \frac{i}{D_{\infty}^2} R_{12}(x)
            \left(\cos \zeta_2(x) J_{\alpha}(n \arccos x) +
            \sin \zeta_2(x) J_{\alpha}'(n \arccos x) \right) \right],
\end{eqnarray}
where $R_{11}(x) = 1 + O(\frac{1}{n})$ and $R_{12}(x) = O(\frac{1}{n})$ as $n \to\infty$,
uniformly for $x \in (1-\delta,1)$, and where
\[
    \zeta_1(x) = \frac{1}{2} \arccos x + \psi(x) + \frac{\alpha\pi}{2},
    \qquad
    \zeta_2(x) = - \frac{1}{2} \arccos x + \psi(x) + \frac{\alpha\pi}{2},
\]
with $\psi$ given by (\ref{functiepsi}).
There is an analogous expression for $x$ in the interval $(-1,-1+\delta)$.
\end{remark}

\section{Proof of Theorem 1.1 and Corollary 1.2}
\setcounter{equation}{0}

In this section we prove Theorem \ref{TheoremUniversalityEdge}
and Corollary \ref{Corollary2UniversalityEdge}. We follow  the
work of Deift et al.\ \cite{Deift,DKMVZ1}. Recall that
\begin{eqnarray} \nonumber
    K_n(x,y) & = & \sqrt{w(x)} \sqrt{w(y)}\, \frac{\gamma_{n-1}}{\gamma_n}\,
    \frac{p_n(x) p_{n-1}(y) - p_{n-1}(x) p_n(y)}{x-y} \\
    & = & \nonumber \sqrt{w(x)}\sqrt{w(y)}\, \gamma_{n-1}^2\,
    \frac{\pi_n(x) \pi_{n-1}(y) - \pi_{n-1}(x) \pi_n(y)}{x-y},
\end{eqnarray}
where $\pi_n$ is the monic orthogonal polynomial of degree $n$ with respect to the modified Jacobi
weight $w$. As in \cite{Deift,DKMVZ1} we replace the polynomials $\pi_{n-1}$ and $\pi_n$ by the
appropriate entries of $Y$, see (\ref{RHPYsolution}), to obtain
\begin{equation} \label{KninY}
    K_n(x,y) = -\frac{1}{2\pi i} \sqrt{w(x)}\sqrt{w(y)}\,
    \frac{Y_{11}(x) Y_{21}(y) - Y_{21}(x) Y_{11}(y)}{x-y}.
\end{equation}
Thus, $K_n$ can be expressed in terms of the first column of $Y$.
The asymptotic behavior of $Y$ follows from the transformations
$Y \mapsto T \mapsto S \mapsto R$ described in section 2, and the
behavior (\ref{asymptoticsR})--(\ref{determinantR}) of $R$.

\subsection{Proof of Theorem 1.1 (a)}
We will first express $Y_{11}$ and $Y_{21}$ in terms of $R$. In the following, $\delta > 0$ will be
a small but fixed number. This number is the radius of the disks $U_{\delta}$ and
$\tilde{U}_{\delta}$ used in the local RH analysis around $\pm 1$.

\begin{lemma}\label{ThAsBulk}
    We have for $x\in(-1+\delta,1-\delta)$,
    \begin{equation} \label{columnYinbulk}
        \begin{pmatrix}
            Y_{11}(x) \\
            Y_{21}(x)
        \end{pmatrix}
        =
        \frac{1}{\sqrt{w(x)}}\, 2^{-n\sigma_3} L_+(x)
        \begin{pmatrix}
            e^{i n \arccos x}  \\
            e^{-in \arccos x}
        \end{pmatrix},
    \end{equation}
    with $L(z)$ given by
    \begin{equation}\label{DefinitieL}
        L(z) = R(z)D_\infty^{\sigma_3}
        \begin{pmatrix}
            \frac{a(z) + a(z)^{-1}}{2} & \frac{a(z) - a(z)^{-1}}{2i} \\[1ex]
            \frac{a(z) - a(z)^{-1}}{-2i} & \frac{a(z) + a(z)^{-1}}{2}
        \end{pmatrix}
        e^{i\psi(z)\sigma_3},
    \end{equation}
    where
    \begin{equation}\label{DefinitiePsi}
        \psi(x) = \frac{1}{2}\Bigl(\alpha(\arccos x-\pi)+\beta \arccos x\Bigr)
        +\frac{\sqrt{1-x^2}}{2\pi}\PVint_{-1}^1\frac{\log h(t)}{\sqrt{1-t^2}}
        \frac{dt}{t-x}.
    \end{equation}
    The matrices $L_+(x)$ and $\frac{d}{dx}L_+(x)$ are uniformly bounded
    for $x \in (-1+\delta, 1-\delta)$ as $n\to\infty$, and
    \begin{equation} \label{determinantL+}
        \det L_+(x) \equiv 1, \qquad \mbox{for } x \in (-1+\delta,1-\delta).
    \end{equation}
\end{lemma}
\begin{remark}
The singular integral (\ref{DefinitiePsi}) is being understood in the sense of
the principal value. It may be shown that $D_+(x) = \sqrt{w(x)} \exp(-i\psi(x))$ so that
$-\psi(x)$ is the argument of the Szeg\H{o} function on the interval.
\end{remark}
\begin{proof}
    We use the series of transformations $Y \mapsto T \mapsto S \mapsto R$
    and we unfold them for $z$ in the upper part of the lens but outside the disks
    $U_\delta$ and $\tilde{U}_\delta$, and then take the limit  to the interval
    $(-1+\delta,1-\delta)$. Thus, let $z$ be in the upper part of the lens but
    outside the disks $U_\delta$ and
    $\tilde{U}_\delta$. We then have by (\ref{YinT}), (\ref{TinS}) and (\ref{DefRBuiten})
    \begin{equation}\label{Bulk1}
            Y(z) =
            2^{-n\sigma_3}R(z)N(z)
            \begin{pmatrix}
                \varphi(z)^n & 0 \\
                w(z)^{-1}\varphi(z)^{-n} & \varphi(z)^{-n}
            \end{pmatrix}.
    \end{equation}
    Inserting the expression (\ref{RHPNsolution}) for $N$ into (\ref{Bulk1}), we
    obtain  for the first column of $Y$
    \begin{equation}\label{PrThAsBulkEq}
        \begin{pmatrix}
            Y_{11}(z) \\
            Y_{21}(z)
        \end{pmatrix}
        =
        2^{-n\sigma_3}R(z)D_\infty^{\sigma_3}
        \begin{pmatrix}
            \frac{a(z) + a(z)^{-1}}{2} & \frac{a(z) - a(z)^{-1}}{2i} \\[1ex]
            \frac{a(z) - a(z)^{-1}}{-2i} & \frac{a(z) + a(z)^{-1}}{2}
        \end{pmatrix}
        \begin{pmatrix}
            \frac{1}{D(z)}\, \varphi(z)^{n} \\[2ex]
            \frac{D(z)}{w(z)}\, \varphi(z)^{-n}
        \end{pmatrix}.
    \end{equation}
    We now take the limit $z\to x\in(-1+\delta,1-\delta)$.
    Since $D_+(x)=\sqrt{w(x)}e^{-i\psi(x)}$ and $\varphi_+(x) = \exp(i \arccos x)$,
    (\ref{columnYinbulk}) now follows from (\ref{PrThAsBulkEq}).

    Note that $R(x)$ and $\frac{d}{dx}R(x)$ are uniformly bounded for $x\in
    (-1+\delta,1-\delta)$ as $n\to\infty$.
    Let $U$ be a neighborhood of $[-1,1]$ such that $\log h$ is defined
    and analytic in $U$, and let $\gamma$ be a closed contour in $U \setminus [-1,1]$
    encircling the interval $[-1,1]$ once in the positive direction. Via
    contour deformation, we may write $\psi$ in the form
    \[
        \psi(x) = \frac{1}{2}\Bigl(\alpha(\arccos x-\pi)+\beta \arccos x\Bigr)
        +\frac{\sqrt{1-x^2}}{4\pi i} \oint_{\gamma} \frac{\log h(\zeta)}{(\zeta^2-1)^{1/2}}
        \frac{d\zeta}{\zeta-x}.
    \]
    Thus $\psi$ has an analytic extension to a neighborhood of $(-1,1)$. This implies
    that $\psi$ and its derivative are bounded on $(-1+\delta, 1-\delta)$.
     From the explicit form (\ref{DefinitieL}) of $L_+$
    we then find that $L_+(x)$ and $\frac{d}{dx}L_+(x)$ are uniformly
    bounded for $x\in (-1+\delta,1-\delta)$ as $n\to\infty$.
    Since $\det R(z)\equiv 1$ it is easy to see from (\ref{DefinitieL})
    that $\det L_+(x)\equiv 1$.
\end{proof}

\begin{varproof}\textbf{of Theorem 1.1 (a)}
    Letting $y \to x$ in (\ref{KninY}) we get
    \begin{eqnarray}
        \nonumber
            K_n(x,x) & = &
            \frac{1}{2\pi i} w(x) \left( Y_{11}(x) Y'_{21}(x) - Y_{21}(x) Y'_{11}(x)\right) \\[1ex]
        &=& \nonumber
            \frac{1}{2\pi i}\det
            \begin{pmatrix}
                \sqrt{w(x)} Y_{11}(x) & \sqrt{w(x)}Y'_{11}(x) \\
                \sqrt{w(x)} Y_{21}(x) & \sqrt{w(x)}Y'_{21}(x)
            \end{pmatrix} \\[1ex]
        &=&\label{PrlDeq1}
            \frac{1}{2\pi i}\det
            \begin{pmatrix}
                2^n \sqrt{w(x)}Y_{11}(x) & \frac{d}{dx}\left(2^n \sqrt{w(x)}Y_{11}(x)\right)  \\
                2^{-n} \sqrt{w(x)}Y_{21}(x) & \frac{d}{dx}\left(2^{-n} \sqrt{w(x)}Y_{21}(x)\right)
            \end{pmatrix}.
    \end{eqnarray}
    By (\ref{columnYinbulk}) the matrix in (\ref{PrlDeq1}) is equal to
    \begin{eqnarray*}
    \lefteqn{
        L_+(x)
        \begin{pmatrix}
            e^{in \arccos x} & -\frac{in}{\sqrt{1-x^2}} e^{in \arccos x} \\
            e^{-in \arccos x} & \frac{in}{\sqrt{1-x^2}} e^{-in \arccos x}
        \end{pmatrix}
        +  \left(\frac{d}{dx} L_+(x)\right)
        \begin{pmatrix}
            0 & e^{in \arccos x} \\
            0 & e^{-in \arccos x}
        \end{pmatrix}
    } \\[1ex]
    & & =
        L_+(x) e^{in \arccos x \sigma_3}
        \begin{pmatrix}
            1 & -\frac{in}{\sqrt{1-x^2}} \\
            1 & \frac{in}{\sqrt{1-x^2}}
        \end{pmatrix}
        + \left(\frac{d}{dx} L_+(x)\right)
        \begin{pmatrix}
            0 & e^{in \arccos x} \\
            0 & e^{-in \arccos x}
        \end{pmatrix}.
    \end{eqnarray*}
    Since $L_+(x)$ and $e^{in \arccos x \sigma_3}$ have determinant one, it follows that
    \begin{equation} \label{PrlDeq2}
        K_n(x,x) = \frac{1}{2\pi i} \det
        \left[
        \begin{pmatrix}
            1 & -\frac{in}{\sqrt{1-x^2}} \\
            1 & \frac{in}{\sqrt{1-x^2}}
        \end{pmatrix}
        + e^{-in \arccos x \sigma_3} L_+^{-1}(x) \left(\frac{d}{dx} L_+(x)\right)
        \begin{pmatrix}
            0 & e^{in \arccos x} \\
            0 & e^{-in \arccos x}
        \end{pmatrix}
        \right].
    \end{equation}
    The entries of $L_+(x)$ and $\frac{d}{dx} L_+(x)$ are uniformly bounded
    by Lemma \ref{ThAsBulk}. Since $\det L_+(x) = 1$, also the entries of
    $L_+^{-1}(x)$ are uniformly bounded.
    Thus, we have uniformly for $x \in (-1+\delta,1-\delta)$,
    \begin{eqnarray}
        \nonumber
        K_n(x,x) &=&
        \frac{1}{2\pi i} \det
        \begin{pmatrix}
            1 & -\frac{in}{\sqrt{1-x^2}} + O(1) \\
            1 & \frac{in}{\sqrt{1-x^2}} + O(1)
        \end{pmatrix}
        =
        \frac{1}{2\pi i} \left( \frac{2in}{\sqrt{1-x^2}} + O(1) \right) \\[1ex]
        \nonumber
        & = &
            \frac{n}{\pi \sqrt{1-x^2}} + O(1).
    \end{eqnarray}
    This proves part (a) of Theorem 1.1.
\end{varproof}

\subsection{Proof of Theorem 1.1 (b)}

\begin{varproof}\textbf{of Theorem  1.1 (b)}
    Let $x \in (-1,1)$ and $u, v \in \mathbb R$.
    For the sake of brevity, we use  $u_{x,n}$ and $v_{x,n}$ to denote
    $x+\frac{u}{n\xi(x)}$ and $x+\frac{v}{n\xi(x)}$, respectively.
    We write
    \[
        \widehat{K}_n(u,v) = \frac{1}{n \xi(x)} K_n ( u_{x,n}, v_{x,n}).
    \]
    From (\ref{KninY}) we then have
    \begin{eqnarray} \nonumber
        \widehat{K}_n(u,v) &=&
            -\frac{1}{2\pi i(u-v)}  \sqrt{w(u_{x,n})}  \sqrt{w(v_{x,n})}
            \det
            \begin{pmatrix}
                Y_{11}(u_{x,n}) & Y_{11}(v_{x,n}) \\
                Y_{21}(u_{x,n}) & Y_{21}(v_{x,n})
            \end{pmatrix} \\[1ex]
        & = & \label{PrThBulkEq1a}
            -\frac{1}{2\pi i(u-v)}
            \det
            \begin{pmatrix}
                2^n \sqrt{w(u_{x,n})} Y_{11}(u_{x,n}) & 2^n \sqrt{w(v_{x,n})} Y_{11}(v_{x,n}) \\
                2^{-n} \sqrt{w(u_{x,n})} Y_{21}(u_{x,n}) & 2^{-n} \sqrt{w(v_{x,n})} Y_{21}(v_{x,n})
            \end{pmatrix}.
    \end{eqnarray}
    Next, we use the expression (\ref{columnYinbulk}) for $Y_{11}$ and $Y_{21}$
    in (\ref{PrThBulkEq1a}) to obtain
    \begin{eqnarray}
        \nonumber
        \widehat{K}_n(u,v) &=&
            -\frac{1}{2\pi i(u-v)} \det\left[
            L_+(u_{x,n})
            \begin{pmatrix} e^{in\arccos u_{x,n}} & 0 \\ e^{-in\arccos u_{x,n}} & 0
            \end{pmatrix} \right. \\[1ex]
        & & \nonumber
            + \, \left.
            L_+(v_{x,n})
            \begin{pmatrix} 0 & e^{in\arccos v_{x,n}} \\ 0 & e^{-in\arccos v_{x,n}}
            \end{pmatrix} \right] \\[2ex]
        & = & \nonumber
            -\frac{1}{2\pi i(u-v)} \det\left[
            L_+(v_{x,n})
            \begin{pmatrix}
                e^{in \arccos u_{x,n}} & e^{in \arccos v_{x,n}} \\
                e^{-in \arccos u_{x,n}} & e^{-in \arccos v_{x,n}}
            \end{pmatrix}
            \right. \\[1ex]
        \label{PrThBulkEq1}
        & &
            \left.
            +\,
            \left[L_+(u_{x,n})-L_+(v_{x,n})\right]
            \begin{pmatrix}
                e^{i n \arccos u_{x,n}} & 0 \\
                e^{-in \arccos u_{x,n}} & 0
            \end{pmatrix}
            \right].
    \end{eqnarray}
    Since $\det L_+(v_{x,n}) = 1$, we then get
    \begin{eqnarray}
        \nonumber
            \widehat{K}_n(u,v) & = &
            -\frac{1}{2\pi i(u-v)} \det\left[
            \begin{pmatrix}
                e^{in \arccos u_{x,n}} & e^{in \arccos v_{x,n}} \\
                e^{-in \arccos u_{x,n}} & e^{-in \arccos v_{x,n}}
            \end{pmatrix}
            \right. \\[1ex]
            \label{PrThBulkEq1b}
        & &
            \left.
            +\, L_+^{-1}(v_{x,n})
            \left[L_+(u_{x,n})-L_+(v_{x,n})\right]
            \begin{pmatrix}
                e^{i n \arccos u_{x,n}} & 0 \\
                e^{-in \arccos u_{x,n}} & 0
            \end{pmatrix}
            \right].
    \end{eqnarray}
    Since $\frac{d}{dx}L_+(x)$ is uniformly bounded as $n\to\infty$,
    we have by the mean value theorem
    \[
        L_+(u_{x,n}) - L_+(v_{x,n}) = O\left(u_{x,n}-v_{x,n}\right) =
        O\left(\frac{u-v}{n}\right),
    \]
    uniformly for $x \in (-1+\delta, 1-\delta)$ and for $u,v$ in compact subsets of $\mathbb R$.
    Since $L_+(x)$ is uniformly bounded, and $\det L_+(x) = 1$, we have
    that $L_+^{-1}(v_{x,n})$ is uniformly bounded as well, so that
    \begin{equation}\label{PrThBulkEq2}
        L_+^{-1}(v_{x,n})
        [L_+(u_{x,n})-L_+(v_{x,n})]
        \begin{pmatrix}
            e^{in\arccos u_{x,n}} & 0 \\
            e^{-in\arccos u_{x,n}} & 0
        \end{pmatrix}
        =
        \begin{pmatrix}
            O\left(\frac{u-v}{n}\right) & 0 \\[1ex]
            O\left(\frac{u-v}{n}\right) & 0
        \end{pmatrix}.
    \end{equation}
    Thus by (\ref{PrThBulkEq1b}) and (\ref{PrThBulkEq2})
    \begin{eqnarray}
        \nonumber
        \widehat{K}_n(u,v) &=&
           - \frac{1}{2\pi i(u-v)}
            \det
            \begin{pmatrix}
                e^{in \arccos u_{x,n}} + O\left(\frac{u-v}{n}\right) & e^{in \arccos v_{x,n}} \\[1ex]
                e^{-in \arccos u_{x,n}} +O\left(\frac{u-v}{n}\right) & e^{-in \arccos v_{x,n}}
            \end{pmatrix} \\[1ex]
        \nonumber
        &=&
            -\frac{1}{2\pi i(u-v)}
            \det
            \begin{pmatrix}
                e^{in \arccos u_{x,n}} & e^{in \arccos v_{x,n}} \\
                e^{-in\arccos u_{x,n}} & e^{-in \arccos v_{x,n}}
            \end{pmatrix}
            +O\left(\frac{1}{n}\right) \\[1ex]
        \nonumber
        & = &
            - \frac{1}{2\pi i(u-v)}
            \left( e^{in (\arccos u_{x,n}-\arccos v_{x,n})} -
                  e^{-in (\arccos u_{x,n}- \arccos v_{x,n})} \right)
            +O\left(\frac{1}{n}\right) \\[1ex]
        & = & - \frac{\sin(n (\arccos u_{x,n} - \arccos v_{x,n}))}{\pi(u-v)}
            + O\left(\frac{1}{n} \right),
    \end{eqnarray}
    uniformly for $x \in (-1+\delta, 1-\delta)$ and $u,v$ in compact
    subsets of $\mathbb{R}$.
    Since
    \[
        n (\arccos u_{x,n}-\arccos v_{x,n}) = -\pi(u-v)(1 + O(1/n)),\qquad \mbox{as $n \to \infty$,}
    \]
    also uniformly for $x \in (-1+\delta, 1-\delta)$ and $u,v$ in compact
    subsets of $\mathbb R$,  part (b) of Theorem 1.1 follows.
\end{varproof}

\subsection{Proof of Theorem 1.1 (c)}

To prove part (c) of Theorem 1.1, we start with a result similar
to Lemma \ref{ThAsBulk}.
\begin{lemma}\label{UniversalityEdgeTheorem1}
    For $x\in (1-\delta,1)$, we have
    \begin{equation}\label{UniversalityEdgeTheorem1Eq}
        \begin{pmatrix}
            Y_{11}(x) \\
            Y_{21}(x)
        \end{pmatrix}
        =
        \sqrt{\frac{2\pi n}{w(x)}}\, 2^{-n\sigma_3}M_+(x)
        \begin{pmatrix}
            J_\alpha(n\arccos x) \\
            \frac{i}{2}\arccos x J'_\alpha(n\arccos x)
        \end{pmatrix},
    \end{equation}
    with $M(z)$ given by
    \begin{equation}\label{DefinitieM}
        M(z)
        =
        R(z)N(z)W(z)^{\sigma_3}\frac{1}{\sqrt{2}}
        \begin{pmatrix}
            1 & -i \\
            -i & 1
        \end{pmatrix}
        f(z)^{\sigma_3/4},
    \end{equation}
    where $R$ is the result of the transformations $Y \mapsto T \mapsto S \mapsto R$
    of the RH problem, the matrix valued function $N$ is given
    by {\rm (\ref{RHPNsolution})}, and the scalar functions $f$ and $W$ are given by {\rm
    (\ref{deff(z)})} and {\rm (\ref{W^2(z)})}, respectively.

    $M$ is analytic in $U_{\delta}$ with $M(z)$ and $\frac{d}{dz}M(z)$ uniformly bounded for $z\in U_\delta$
    as $n\to\infty$. Furthermore, we have
    \[
        \det M(z) \equiv 1.
    \]
\end{lemma}
\begin{proof}
    As in the proof of Lemma \ref{ThAsBulk} we unravel the transformations
    $Y\mapsto T\mapsto S\mapsto R$,
    but now for $z$ in the upper part of the lens and
    inside the disk $U_\delta$. We then have by (\ref{YinT}),
    (\ref{TinS}), (\ref{RHPPsolution}) and (\ref{DefRU})
    \begin{equation} \label{AsymptotiekRond1Eq1}
        Y(z)
        =
        2^{-n\sigma_3}R(z)E_n(z)\Psi\left(n^2 f(z)\right)W(z)^{-\sigma_3}
        \begin{pmatrix}
            1 & 0 \\
            w(z)^{-1} & 1
        \end{pmatrix}.
    \end{equation}
    Since $\Im z>0$, we have by (\ref{W^2(z)}) that $W(z)=w^{1/2}(z)e^{i\pi\alpha/2}$.
    Inserting this into (\ref{AsymptotiekRond1Eq1}) we get for the first column of $Y$
    \begin{equation} \label{AsymptotiekRond1Eq2}
        \begin{pmatrix}
            Y_{11}(z) \\
            Y_{21}(z)
        \end{pmatrix}
        =
        w^{-1/2}(z) 2^{-n\sigma_3} R(z) E_n(z)\Psi\left(n^2f(z)\right)
        \begin{pmatrix}
            e^{-i\pi\alpha/2} \\
            e^{i\pi\alpha/2}
        \end{pmatrix}.
    \end{equation}
    Since $z$ is in the upper part of the lens and inside the disk $U_\delta$, we have
    $2\pi/3<\arg n^2 f(z)<\pi$, see \cite[section 6]{KMVV}, and we thus use (\ref{RHPPSIsolution2}) to
    evaluate $\Psi(n^2 f(z))$. From formulas 9.1.3 and 9.1.4 of \cite{AbramowitzStegun} we then have
    \begin{equation}\label{Pr1}
        \Psi\left(n^2 f(z)\right)
        \begin{pmatrix}
            e^{-i\pi\alpha/2} \\
            e^{i\pi\alpha/2}
        \end{pmatrix}
        =
        \begin{pmatrix}
            J_{\alpha}(2n(-f(z))^{1/2}) \\[1ex]
            2\pi n f^{1/2}(z)J'_{\alpha}(2n(-f(z))^{1/2})
        \end{pmatrix}.
    \end{equation}
    By (\ref{eq536}) and (\ref{DefinitieM}) we have $R(z)E_n(z) = M(z)(2\pi n)^{\sigma_3/2}$.
    Inserting this and (\ref{Pr1}) into (\ref{AsymptotiekRond1Eq2}) we get
    \begin{equation} \label{AsymptotiekRond1Eq3}
        \begin{pmatrix}
            Y_{11}(z) \\
            Y_{21}(z)
        \end{pmatrix}
        =
        \frac{\sqrt{2\pi n}}{w^{1/2}(z)} 2^{-n\sigma_3} M(z)
        \begin{pmatrix}
            J_{\alpha}(2n(-f(z))^{1/2}) \\[1ex]
            f^{1/2}(z)J'_{\alpha}(2n(-f(z))^{1/2})
        \end{pmatrix}.
    \end{equation}
    We now take the limit $z\to x\in(1-\delta,1)$. By (\ref{deff(z)}), and since
    $\varphi_+(x)=\exp(i\arccos x)$ we have
    $f^{1/2}_+(x)=\frac{i}{2} \arccos x$, so that $2(-f_+(x))^{1/2}=\arccos x$.
    Inserting this into (\ref{AsymptotiekRond1Eq3}), we obtain
    (\ref{UniversalityEdgeTheorem1Eq}).

    $M$ is analytic in $U_\delta$ since both $E_n$, see \cite[Proposition 6.5]{KMVV},
    and $R$ are analytic in $U_\delta$. So, we may write $M(x)$
    instead of $M_+(x)$ in (\ref{UniversalityEdgeTheorem1Eq}).

    By (\ref{asymptoticsR}) and (\ref{asymptoticsdR}) we have
    that $R(z)$ and $\frac{d}{dz}R(z)$ are uniformly bounded for $z\in U_\delta$
    as $n\to\infty$. Since
    \[
        N(z)W(z)^{\sigma_3}\frac{1}{\sqrt{2}}
        \begin{pmatrix}
            1 & -i \\
            -i & 1
        \end{pmatrix}
        f(z)^{\sigma_3/4}
    \]
    is analytic for $z\in U_\delta$ and does not depend on $n$, we  have from
    (\ref{DefinitieM}) that also $M(z)$ and
    $\frac{d}{dz}M(z)$ are uniformly bounded for $z\in U_\delta$ as $n\to\infty$.

    The fact that $\det M(z)\equiv 1$ follows easily from (\ref{RHPNsolution}),
    (\ref{determinantR}),  and (\ref{DefinitieM}).
\end{proof}

\begin{remark}
    We check that (\ref{UniversalityEdgeTheorem1Eq}) yields
    $\begin{pmatrix}
        Y_{11}(x) \\
        Y_{21}(x)
    \end{pmatrix}=O(1)$ as $x\nearrow 1$, which is in agreement with the fact
    that  $Y_{11}$ and $Y_{21}$ are polynomials, see (\ref{RHPYsolution}).
    Since $\arccos x=O\left((1-x)^{1/2}\right)$ as $x\nearrow 1$,
    we have by formula 9.1.10 of \cite{AbramowitzStegun}
    \begin{equation}\label{Newremarkedge}
        \begin{pmatrix}
            J_\alpha(n\arccos x) \\
            \arccos x\, J'_\alpha(n\arccos x)
        \end{pmatrix}
        =
        \begin{pmatrix}
            O\left((1-x)^{\alpha/2}\right) \\
            O\left((1-x)^{\alpha/2}\right)
        \end{pmatrix},
        \qquad\mbox{as }x\nearrow 1.
    \end{equation}
    Since $M$  is analytic near 1, see Lemma \ref{UniversalityEdgeTheorem1}, we have $M(x)=O(1)$ as
    $x\nearrow 1$. Inserting (\ref{Newremarkedge}) into
    (\ref{UniversalityEdgeTheorem1Eq}) and noting that $w(x)^{-1} = O((1-x)^{-\alpha})$
    as $x \nearrow 1$, we then have indeed that $Y_{11}(x)$ and $Y_{21}(x)$ remain bounded
    as $x\nearrow 1$.
\end{remark}

We also need the asymptotic behavior of $J_{\alpha}(\tilde{u}_n)$ and
$J'_\alpha(\tilde{u}_n)$ as $n\to\infty$,
where we put $u_n = 1-\frac{u}{2n^2}$ and $\tilde{u}_n = n \arccos u_n$.
These will be contained in the next lemma.

\begin{lemma}\label{UniversalityEdgeLemma1}
    Let $u \in(0,\infty)$, $u_n = 1- \frac{u}{2n^2}$, and $\tilde{u}_n = n \arccos u_n$.
    We then have as $n\to\infty$,
    \begin{eqnarray}
        \label{UniversalityEdgeLemma1Eq1}
        \lefteqn{
            \tilde{u}_n = \sqrt{u} + O\left(\frac{u^{\frac{3}{2}}}{n^2}\right),} \\[1ex]
        \label{UniversalityEdgeLemma1Eq2}
        \lefteqn{
            J_\alpha(\tilde{u}_n) = J_\alpha(\sqrt{u})+O\left(\frac{u^{\frac{\alpha}{2}+1}}{n^2}\right),} \\[1ex]
        \label{UniversalityEdgeLemma1Eq3}
        \lefteqn{
            J'_\alpha(\tilde{u}_n) = J'_\alpha(\sqrt{u})+O\left(\frac{u^{\frac{\alpha}{2}+\frac{1}{2}}}{n^2}\right).}
    \end{eqnarray}
    The error terms hold uniformly for $u$ in bounded subsets of $(0,\infty)$.
\end{lemma}

\begin{proof}
    Since $z^{-1/2} \arccos(1-z)$ is analytic in a neighborhood of $0$ with
    expansion
    \[
        z^{-1/2} \arccos(1-z) = \sqrt{2} +  O(z), \qquad\mbox{as $z\to 0$,}
    \]
    we easily get (\ref{UniversalityEdgeLemma1Eq1}).

    By formula 9.1.10 of \cite{AbramowitzStegun} we know that $J_\alpha(z)=z^{\alpha}G(z)$
    with $G$ an entire function. From (\ref{UniversalityEdgeLemma1Eq1}) and
    Taylor's formula we then get uniformly for $u$ in bounded subsets of
    $(0,\infty)$,
    \begin{eqnarray}
        \nonumber
        J_\alpha(n \arccos u_n)
        &=&
            u^\frac{\alpha}{2}\left(1+O\left(\frac{u}{n^2}\right)\right)\left(G(\sqrt u)+
            O\left(\frac{u^{\frac{3}{2}}}{n^2}\right)\right) \\[1ex]
        \nonumber
        &=&
            J_{\alpha}(\sqrt u)+O\left(\frac{u^{\frac{\alpha}{2}+1}}{n^2}\right),
    \end{eqnarray}
    which is (\ref{UniversalityEdgeLemma1Eq2}).
    The proof of (\ref{UniversalityEdgeLemma1Eq3}) follows in a similar fashion,
    using
    \[
        J'_\alpha(z) = \alpha z^{\alpha-1}G(z)+z^\alpha G'(z),
    \]
    and Taylor's formula again.
\end{proof}

Now we are ready for the proof of our main result.
\begin{varproof}\textbf{of Theorem 1.1 (c)}
    Let $u, v \in (0,\infty)$ and define
    \begin{equation} \label{definitieunvn}
        u_n = 1- \frac{u}{2n^2}, \quad v_n = 1-\frac{v}{2n^2}, \qquad
        \tilde{u}_n = n \arccos u_n, \quad \tilde{v}_n = n \arccos v_n.
    \end{equation}
    We put
    \begin{equation} \label{definitieDn}
        D_n(u,v) = \frac{1}{2n^2} K_n(u_n, v_n).
    \end{equation}
    From (\ref{KninY}) we then have
    \begin{eqnarray}
        \nonumber
        D_n(u,v) & = &
            \frac{1}{2\pi i (u-v)}\sqrt{w(u_n)} \sqrt{w(v_n)}
            \det
            \begin{pmatrix}
                Y_{11}(u_n) & Y_{11}(v_n) \\
                Y_{21}(u_n) & Y_{21}(v_n)
            \end{pmatrix} \\[1ex]
        & = &
            \frac{1}{2\pi i (u-v)}
            \det
            \begin{pmatrix}
                2^n \sqrt{w(u_n)} Y_{11}(u_n) & 2^n \sqrt{w(v_n)} Y_{11}(v_n) \\
                2^{-n} \sqrt{w(u_n)} Y_{21}(u_n) & 2^{-n} \sqrt{w(v_n)} Y_{21}(v_n)
            \end{pmatrix}.
        \label{PrTh1Eq1a}
    \end{eqnarray}
    Next, we replace the two columns in the determinant in (\ref{PrTh1Eq1a})
    by the expression (\ref{UniversalityEdgeTheorem1Eq}) we found
    in Lemma \ref{UniversalityEdgeTheorem1}. It follows that
    \begin{equation} \label{PrTh1Eq1b}
            D_n(u,v) =
            \frac{2\pi n}{2\pi i(u-v)}
            \det \left[ M(u_n)
            \begin{pmatrix}
                J_\alpha(\tilde{u}_n) & 0 \\[1ex]
                \frac{i}{2n} \tilde{u}_n J'_\alpha(\tilde{u}_n) & 0
            \end{pmatrix}
                +M(v_n)
            \begin{pmatrix}
                0 & J_\alpha(\tilde{v}_n) \\[1ex]
                0 & \frac{i}{2n} \tilde{v}_n J'_\alpha(\tilde{v}_n)
            \end{pmatrix}
            \right]. \qquad
    \end{equation}
    We rewrite the matrix appearing in the determinant in (\ref{PrTh1Eq1b})
    as
    \[
        M(v_n) \left[
        \begin{pmatrix}
            J_{\alpha}(\tilde{u}_n) & J_{\alpha}(\tilde{v}_n) \\[1ex]
            \frac{i}{2n}  \tilde{u}_n J'_\alpha(\tilde{u}_n) &
                \frac{i}{2n} \tilde{v}_n J'_\alpha(\tilde{v}_n)
        \end{pmatrix}
        + M(v_n)^{-1} \bigl[M(u_n)-M(v_n) \bigr]
        \begin{pmatrix}
            J_{\alpha}(\tilde{u}_n) & 0  \\[1ex]
            \frac{i}{2n}  \tilde{u}_n J'_\alpha(\tilde{u}_n) & 0
        \end{pmatrix} \right].
    \]
    Now we use  $\det M(v_n) = 1$ and the fact that $M(z)$ is uniformly bounded
    for $z \in U_{\delta}$, see Lemma \ref{UniversalityEdgeTheorem1},
    to conclude that the entries of $M(v_n)^{-1}$ are uniformly bounded.
    By Lemma \ref{UniversalityEdgeTheorem1}, we also have that $\frac{d}{dz} M(z)$ is uniformly
    bounded so that $M(u_n) - M(v_n) = O\left(\frac{u-v}{n^2}\right)$.
    From Lemma \ref{UniversalityEdgeLemma1} it follows that
    $J_\alpha(\tilde{u}_n)=O(u^{\alpha/2})$ and $\tilde{u}_nJ'_\alpha(\tilde{u}_n)
    =O(u^{\alpha/2})$ uniformly for $u$ in bounded subsets of $(0,\infty)$ as
    $n\to\infty$. Hence we have, uniformly for $u,v$ in bounded subsets of $(0,\infty)$,
    \begin{equation}\label{Kansintervalafschattingeq2}
        M(v_n)^{-1} \bigl[M(u_n)-M(v_n)\bigr]
        \begin{pmatrix}
            J_\alpha(\tilde{u}_n) & 0 \\[1ex]
            \frac{i}{2n} \tilde{u}_n J'_\alpha(\tilde{u}_n ) & 0
        \end{pmatrix}
        =
        \begin{pmatrix}
            O\left(\frac{u-v}{n^2}\, u^{\frac{\alpha}{2}}\right) & 0 \\[1ex]
            O\left(\frac{u-v}{n^2}\, u^{\frac{\alpha}{2}}\right) & 0
        \end{pmatrix}.
    \end{equation}
    It now follows that (we use $\det M(v_n) = 1$)
    \begin{equation}
        D_n(u,v) =
            \frac{1}{2 (u-v)}
            \det
            \begin{pmatrix}
                J_\alpha( \tilde{u}_n)+O\left(\frac{u-v}{n^2}\, u^{\frac{\alpha}{2}}\right) &
                J_\alpha( \tilde{v}_n) \\[1ex]
                \tilde{u}_n J'_\alpha( \tilde{u}_n)+O\left(\frac{u-v}{n}\, u^{\frac{\alpha}{2}}\right) &
               \tilde{v}_n J'_\alpha(\tilde{v}_n)
            \end{pmatrix}.
    \end{equation}
    Since $J_\alpha(\tilde{v}_n)=O(v^{\alpha/2})$ and $\tilde{v}_n J'_\alpha(\tilde{v}_n)=O(v^{\alpha/2})$
    as $n\to\infty$, we then get  uniformly for $u,v$ in bounded subsets of $(0,\infty)$,
    \begin{equation} \label{Kansintervalafschattingeq3}
            D_n(u,v) =
            \frac{1}{2(u-v)}
            \det
            \begin{pmatrix}
                J_\alpha(\tilde{u}_n) & J_\alpha(\tilde{v}_n) \\[1ex]
                \tilde{u}_n J'_\alpha(\tilde{u}_n) & \tilde{v}_n J'_\alpha(\tilde{v}_n)
            \end{pmatrix}
            + O\left(\frac{u^\frac{\alpha}{2}v^\frac{\alpha}{2}}{n}\right).
    \end{equation}
    In the determinant in (\ref{Kansintervalafschattingeq3}) we can
    replace $\tilde{u}_n$ and $\tilde{v}_n$ by $\sqrt{u}$ and $\sqrt{v}$
    respectively, and make an error which we could estimate using
    Lemma \ref{UniversalityEdgeLemma1}. However, this estimate would not be
    uniform for $u - v$ close to zero. So we will  be more careful.
    We bring in a factor $u^{-\frac{\alpha}{2}}$ into the first column of
    the determinant in (\ref{Kansintervalafschattingeq3}) and a factor
    $v^{-\frac{\alpha}{2}}$ into the second. Then we subtract the second
    column from the first to obtain
    \begin{equation} \label{Kansintervalafschattingeq4}
            D_n(u,v)
            =
            \frac{u^{\frac{\alpha}{2}}v^{\frac{\alpha}{2}}}{2(u-v)}
            \det
            \begin{pmatrix}
                u^{-\frac{\alpha}{2}}J_\alpha(\tilde{u}_n)-v^{-\frac{\alpha}{2}}J_\alpha(\tilde{v}_n) &
                    v^{-\frac{\alpha}{2}}J_\alpha(\tilde{v}_n) \\[1ex]
                u^{-\frac{\alpha}{2}}\tilde{u}_n J'_\alpha(\tilde{u}_n)-
                    v^{-\frac{\alpha}{2}}\tilde{v}_n J'_\alpha(\tilde{v}_n) &
                    v^{-\frac{\alpha}{2}}\tilde{v}_n J'_\alpha(\tilde{v}_n)
            \end{pmatrix}
            + O\left(\frac{u^\frac{\alpha}{2}v^\frac{\alpha}{2}}{n}\right).\ \
    \end{equation}
    From Lemma  \ref{UniversalityEdgeLemma1} it follows that
    uniformly  for $x$ in bounded subsets of $(0,\infty)$,
    \[
        \frac{d}{dx} \left(x^{-\frac{\alpha}{2}}J_\alpha(n \arccos x_n)
        -x^{-\frac{\alpha}{2}} J_\alpha(\sqrt x)\right)
        =
        O\left(\frac{1}{n^2}\right).
    \]
    Then it easily follows that the $1,1$-entry in the determinant
    in (\ref{Kansintervalafschattingeq4}) is equal to
    \[
        u^{-\frac{\alpha}{2}}J_\alpha(\sqrt u)-v^{-\frac{\alpha}{2}}J_\alpha(\sqrt v)
        + O\left(\frac{u-v}{n^2}\right).
    \]
    Similarly, if we use
    \[
        \frac{d}{dx} \left(x^{-\frac{\alpha}{2}}n \arccos x_nJ'_\alpha(n \arccos x_n)-
        x^{-\frac{\alpha}{2}}\sqrt{x}J'_\alpha(\sqrt x)\right) = O\left(\frac{1}{n^2}\right),
    \]
    we find that the $2,1$-entry is
    \[
        u^{-\frac{\alpha}{2}}\sqrt u J'_\alpha(\sqrt u)- v^{-\frac{\alpha}{2}}\sqrt v J'_\alpha(\sqrt v)
        + O\left(\frac{u-v}{n^2}\right).
    \]
    From Lemma \ref{UniversalityEdgeLemma1} it also follows that
    we may replace $\tilde{v}_n$ by $\sqrt{v}$ in the second column
    at the expense of an error term $O(\frac{1}{n^2})$. Therefore,
    uniformly for $u,v$ in bounded subsets of $(0,\infty)$,
    \begin{eqnarray}
        \nonumber
        \lefteqn{
            D_n(u,v) = \frac{u^{\frac{\alpha}{2}}v^{\frac{\alpha}{2}}}{2(u-v)} } \\[1ex]
        \nonumber
        & &
            \, \times \, \det
            \begin{pmatrix}
                u^{-\frac{\alpha}{2}}J_\alpha(\sqrt u)-v^{-\frac{\alpha}{2}}J_\alpha(\sqrt v)
                    + O\left(\frac{u-v}{n^2}\right)
                & v^{-\frac{\alpha}{2}}J_\alpha(\sqrt v)+O\left(\frac{1}{n^2}\right) \\[1ex]
                u^{-\frac{\alpha}{2}}\sqrt u J'_\alpha(\sqrt u)- v^{-\frac{\alpha}{2}}\sqrt v J'_\alpha(\sqrt v)
                    + O\left(\frac{u-v}{n^2}\right)
                & v^{-\frac{\alpha}{2}}\sqrt v J'_\alpha(\sqrt v)+O\left(\frac{1}{n^2}\right)
            \end{pmatrix} \\[1ex]
        & & \nonumber
            \qquad + \, O\left(\frac{u^{\frac{\alpha}{2}} v^{\frac{\alpha}{2}}}{n} \right) \\[2ex]
        & & \nonumber
            = \mathbb{J}_\alpha(u,v)+
            \frac{u^{\frac{\alpha}{2}}v^{\frac{\alpha}{2}}}{2(u-v)}
            \det
            \begin{pmatrix}
                u^{-\frac{\alpha}{2}}J_\alpha(\sqrt u)-v^{-\frac{\alpha}{2}}J_\alpha(\sqrt v)
                & O\left(\frac{1}{n^2}\right) \\[1ex]
                u^{-\frac{\alpha}{2}}\sqrt u J'_\alpha(\sqrt u)- v^{-\frac{\alpha}{2}}\sqrt v J'_\alpha(\sqrt v)
                & O\left(\frac{1}{n^2}\right)
            \end{pmatrix} \\[1ex]
        & & \label{Kansintervalafschattingeq6}
            \qquad +\,
            O\left(\frac{u^\frac{\alpha}{2}v^\frac{\alpha}{2}}{n}\right).
    \end{eqnarray}
    Since $z^{-\alpha/2}J_\alpha(\sqrt z)$ is an entire function we get by
    the mean value theorem that
    \[
        \frac{u^{-\frac{\alpha}{2}}J_\alpha(\sqrt u)-v^{-\frac{\alpha}{2}}J_\alpha(\sqrt v)}{u-v}
    \]
    is bounded for $u,v$ in bounded subsets of $(0,\infty)$, and similarly,
    that
    \[
        \frac{u^{-\frac{\alpha}{2}}\sqrt u J'_\alpha(\sqrt u)- v^{-\frac{\alpha}{2}}\sqrt v J'_\alpha(\sqrt
        v)}{u-v}
    \]
    is bounded for $u,v$ in bounded subsets of $(0,\infty)$. Therefore, we have by
    (\ref{Kansintervalafschattingeq6})
    \[
        D_n(u,v) =
        \mathbb{J}_\alpha(u,v)+O\left(\frac{u^\frac{\alpha}{2}v^\frac{\alpha}{2}}{n}\right),
    \]
    uniformly for $u,v$ in bounded subsets of $(0,\infty)$, which completes the
    proof of part (c) of Theorem 1.1.
\end{varproof}

\subsection{Proof of Corollary 1.2}

To prove Corollary \ref{Corollary2UniversalityEdge} we first need
two lemmas.
\begin{lemma}
    For $u,v\in(0,\infty)$ we have as $n \to \infty$,
    \begin{equation}\label{Kansintervalgedrag}
       \frac{1}{2n^2} K_n \left(1-\frac{u}{2n^2}, 1- \frac{v}{2n^2} \right)
       = O\left(u^\frac{\alpha}{2}v^\frac{\alpha}{2}\right).
    \end{equation}
    The error term is uniform for $u,v$ in bounded subsets of $(0,\infty)$.
\end{lemma}

\begin{proof}
    Since
    \[
        \mathbb{J}_\alpha(u,v) = \frac{u^\frac{\alpha}{2}v^\frac{\alpha}{2}}{2(u-v)}\det
        \begin{pmatrix}
            u^{-\frac{\alpha}{2}}J_\alpha(\sqrt u)-v^{-\frac{\alpha}{2}}J_\alpha(\sqrt v)
                & v^{-\frac{\alpha}{2}}J_\alpha(\sqrt v) \\[1ex]
            u^{-\frac{\alpha}{2}}\sqrt u J'_\alpha(\sqrt u)- v^{-\frac{\alpha}{2}}\sqrt v J'_\alpha(\sqrt v)
                 & v^{-\frac{\alpha}{2}}\sqrt v J'_\alpha(\sqrt v)
        \end{pmatrix},
    \]
    and since $v^{-\alpha/2}J_\alpha(\sqrt v)$ and $v^{-\alpha/2}\sqrt v J'_\alpha(\sqrt v)$
    are entire functions of $v$, it is easy to see,
    by the discussion following (\ref{Kansintervalafschattingeq6}), that
    for every $R > 0$, there exists a constant
    $c>0$ so that
    \begin{equation}\label{Jalphaafschatting}
        \mathbb{J}_\alpha(u,v) \leq c u^\frac{\alpha}{2}v^\frac{\alpha}{2},
            \qquad \mbox{for } u,v \in (0,R).
    \end{equation}
    Therefore, (\ref{Kansintervalgedrag}) is an immediate consequence of Theorem
    \ref{TheoremUniversalityEdge}.
\end{proof}

In the second lemma, we let
$D_{n,s}$ and $\mathbb{J}_{\alpha,s}$ be the integral operators with kernels
$D_n(u,v) = \frac{1}{2n^2}K_n(1-\frac{u}{2n^2}, 1- \frac{v}{2n^2})$ and
$\mathbb{J}_\alpha(u,v)$ respectively, acting on $L^2(0,s)$.

\begin{lemma}
    $D_{n,s}$ and $\mathbb J_{\alpha,s}$ are positive trace class operators on $L^2(0,s)$.
\end{lemma}

\begin{proof}
    Let $u_n = 1 - \frac{u}{2n^2}$ and $v_n = 1 - \frac{v}{2n^2}$.
    Since
    \begin{equation}\label{herhalingdefinitieD_n}
        D_n(u,v) = \frac{1}{2 n^2} \sqrt{w(u_n)}\sqrt{w(v_n)}\sum_{j=0}^{n-1} p_j(u_n)p_j(v_n),
    \end{equation}
    we have that $D_{n,s}$ is a finite rank operator and hence a trace class operator.
    Now, for every $f\in L^2(0,s)$ we have by (\ref{herhalingdefinitieD_n})
    \begin{equation} \label{D_npositief}
        \int_0^s \int_0^s D_n(u,v)f(u)f(v)dudv
        = \frac{1}{2n^2} \sum_{j=0}^{n-1}
        \left(\int_0^s\sqrt{w(u_n)}p_j(u_n)f(u) du \right)^2 \geq 0,
    \end{equation}
    so that $D_{n,s}$ is a positive operator.
    Letting $n \to \infty$, we have that
    $D_n(u,v)\to\mathbb{J}_\alpha(u,v)$ for every $u,v$ by Theorem
    \ref{TheoremUniversalityEdge}(c).
    By (\ref{Kansintervalgedrag}), there  is a constant $c>0$ independent of $n$
    so that for $u,v\in(0,s]$,
    \[
        |D_n(u,v)f(u)f(v)| \leq  c u^\frac{\alpha}{2} v^\frac{\alpha}{2} |f(u)||f(v)|.
    \]
    Then by the dominated convergence theorem and (\ref{D_npositief}),
    \begin{equation}\label{Jalphapositief}
        \int_0^s \int_0^s \mathbb{J}_\alpha(u,v)f(u)f(v)dudv
        =
        \lim_{n\to\infty} \int_0^s \int_0^s D_n(u,v)f(u)f(v)dudv \geq 0,
    \end{equation}
    so that $\mathbb J_{\alpha}$ is positive.
    Since $\left|\mathbb{J}_\alpha(u,v)\right| \leq c u^{\alpha/2} v^{\alpha/2}$
    for $u,v\in(0,s]$ we also have
    \[
        |\tr \mathbb{J}_{\alpha,s}| = \left|\int_0^s\mathbb{J}_\alpha(u,u)du\right|
        \leq c\int_0^s u^{\alpha}du < \infty,
    \]
    which implies that $\mathbb{J}_{\alpha,s}$ is a trace class operator.
\end{proof}

\begin{varproof}\textbf{of Corollary \ref{Corollary2UniversalityEdge}}
    It is well known that, see for example \cite{Deift,DKMVZ1},
    \[
        P_n\left(1-\frac{s}{2n^2},1\right) = \sum_{j=0}^{n}\frac{(-1)^j}{j!}\int_0^s \ldots
        \int_0^s \det\left(D_n(u_i,u_k)\right)_{1\leq i,k \leq j}du_1\ldots du_j.
    \]
    For $\alpha \geq 0$, we have that $D_n(u,v)$ is continuous on $[0,s] \times [0,s]$.
    Then it follows as in \cite{Simon} that $P_n(1-\frac{s}{2n^2}, 1)$
    is equal to the Fredholm determinant
    \begin{equation} \label{Pn=Fredholmdeterminant}
        P_n\left(1-\frac{s}{2n^2},1\right) = \det(I-D_{n,s}).
    \end{equation}
    For $\alpha < 0$, the integral kernel $D_n(u,v)$ is not continuous, but
    satisfies an estimate $|D_n(u,v)|\leq c u^{\alpha/2}v^{\alpha/2}$ for $u,v\in (0,s]$,
    for some constant $c>0$. This is enough to establish (\ref{Pn=Fredholmdeterminant})
    also in this case.
    By Theorem 3.4 of \cite{Simon} we have
    \[
        \left|\det(I-D_{n,s})-\det(I-\mathbb{J}_{\alpha,s})\right| \leq
        \|D_{n,s}-\mathbb{J}_{\alpha,s}\norm1
        \exp\left(\|D_{n,s}-\mathbb{J}_{\alpha,s}\norm1 + 2\|\mathbb{J}_{\alpha,s}\norm1+1\right),
    \]
    where $\|\cdot\norm1$ is the trace norm in $L^2(0,s)$. So, in order to prove that
    $\det(I-D_{n,s})$ converges to $\det(I-\mathbb{J}_{\alpha,s})$ as $n\to\infty$, it
    is enough to show that  $D_{n,s}$ tends to $\mathbb{J}_{\alpha,s}$ in trace norm.
    By Theorem 2.20 of \cite{Simon} and the positivity of $D_{n,s}$ and
    $\mathbb{J}_{\alpha,s}$, it then suffices to prove that
    $D_{n,s}\to\mathbb{J}_{\alpha,s}$ weakly, and that
    $\tr D_{n,s} \to \tr \mathbb{J}_{\alpha,s}$ as $n\to\infty$.
    So we have to prove that for $f,g \in L^2(0,s)$,
    \begin{equation} \label{weakconvergence}
        \lim_{n \to \infty} \int_0^s\int_0^s D_n(u,v) f(u) g(v) du dv
        = \int_0^s \int_0^s \mathbb J_{\alpha}(u,v) f(u) g(v) du dv,
    \end{equation}
    and that
    \begin{equation} \label{traceconvergence}
        \lim_{n \to \infty} \int_0^s D_n(u,u) du = \int_0^s \mathbb J_{\alpha}(u,u) du.
    \end{equation}
    Both (\ref{weakconvergence}) and (\ref{traceconvergence}) follow easily
    from the pointwise convergence $D_n(u,v) \to J_{\alpha}(u,v)$, the
    uniform bound $|D_n(u,v)| \leq c u^{\alpha/2} v^{\alpha/2}$ and the
    dominated convergence theorem.
\end{varproof}

\subsection*{Acknowledgements}
    We thank Ken McLaughlin and Walter Van Assche for useful discussions.

\end{document}